\documentclass[useAMS,usenatbib,A4]{mn2e}
\usepackage{graphicx,subfigure,amsmath, amsfonts, amssymb,aas_macros,times,mathrsfs,epstopdf}
\usepackage[usenames,dvipsnames]{color}
\usepackage{helvet}
\usepackage{epsfig,natbib}

\usepackage{adjustbox}
\newcommand{\atlas}{ATLAS$^{\rm 3D}$}
\newcommand{\rco}{$\mathscr{R}_{13}$}
\newcommand{\thco}{$^{13}$CO}
\newcommand{\twco}{$^{12}$CO}

\title[$^{13}$CO(1--0) in ETGs]{Evidence of boosted \thco/\twco\ ratio in early-type galaxies in dense environments}
\author[K. Alatalo et al.]{Katherine Alatalo,$^{1}$\thanks{E-mail:
kalatalo@caltech.edu} Alison F. Crocker,$^{2,3}$ Susanne Aalto,$^{4}$ Timothy A. Davis,$^{5,6}$
\newauthor Kristina Nyland,$^{7}$ Martin Bureau,$^{8}$ Pierre-Alain Duc,$^{9}$ Davor Krajnovi\'c,$^{10}$ 
\newauthor and Lisa M. Young$^{11}$\\
$^{1}$Infrared Processing and Analysis Center, California Institute of Technology, Pasadena, California 91125, USA\\
$^{2}$Department of Physics and Astronomy, University of Toledo, Toledo, OH 43606, USA\\
$^{3}$Department of Physics, Reed College, Portland, Oregon 97202\\
$^{4}$Department of Earth and Space Sciences, Chalmers University of Technology, Onsala Observatory, 439 94 Onsala, Sweden\\
$^{5}$European Southern Observatory, Karl-Schwarzschild-Str. 2, 85748 Garching, Germany\\
$^{6}$Centre for Astrophysics Research, University of Hertfordshire, Hatfield, Herts AL1 9AB, UK\\
$^{7}$Netherlands Institute for Radio Astronomy (ASTRON), Postbus 2, 7990 AA Dwingeloo, The Netherlands\\
$^{8}$Sub-Department of Astrophysics, Department of Physics, University of Oxford, Denys Wilkinson Building, Keble Road, Oxford, OX1 3RH, UK\\
$^{9}$Laboratoire AIM Paris-Saclay, CEA/IRFU/SAp -- CNRS -- Universit\'e Paris Diderot, 91191 Gif-sur-Yvette Cedex, France\\
$^{10}$Leibniz-Institut f\"ur Astrophysik Potsdam (AIP), An der Sternwarte 16, D-14482 Potsdam, Germany\\
$^{11}$Physics Department, New Mexico Institute of Mining and Technology, Socorro, NM 87801, USA\\
}
\begin{document}

\date{Received -- October 8, 2014 -- Accepted -- April 8, 2015}

\pagerange{\pageref{firstpage}--\pageref{lastpage}} \pubyear{2015}

\maketitle

\label{firstpage}

\begin{abstract}
We present observations of \thco(1--0) in 17 Combined Array for Research in Millimeter Astronomy (CARMA) \atlas\ early-type galaxies (ETGs), obtained simultaneously with \twco(1--0) observations.  The \thco\ in six ETGs is sufficiently bright to create images.  In these 6 sources, we do not detect any significant radial gradient in the \thco/\twco\ ratio between the nucleus and the outlying molecular gas.  Using the \twco\ channel maps as 3D masks to stack the \thco\ emission, we are able to detect 15/17 galaxies to $>3\sigma$ (and 12/17 to at least $5\sigma$) significance in a spatially integrated manner.  Overall, ETGs show a wide distribution of \thco/\twco\ ratios, but Virgo cluster and group galaxies preferentially show a \thco/\twco\ ratio about 2 times larger than field galaxies, although this could also be due to a mass dependence, or the CO spatial extent ($R_{\rm CO}/R_{\rm e}$).
ETGs whose gas has a morphologically-settled appearance also show boosted \thco/\twco\ ratios.  We hypothesize that this variation could be caused by (i) the extra enrichment of gas from molecular reprocessing occurring in low-mass stars (boosting the abundance of $^{13}$C to $^{12}$C in the absence of external gas accretion), (ii) much higher pressure being exerted on the midplane gas (by the intracluster medium) in the cluster environment than in isolated galaxies, or (iii) all but the densest molecular gas clumps being stripped as the galaxies fall into the cluster.  Further observations of \thco\ in dense environments, particularly of spirals, as well as studies of other isotopologues, should be able to distinguish between these hypotheses.
\end{abstract}


\section{Introduction}
%%%%%%%%%%  Figure 1  %%%%%%%%%%
\begin{figure*}
\subfigure{\includegraphics[width=0.48\textwidth,clip,trim=2.5cm 0.6cm 1.3cm 1.6cm]{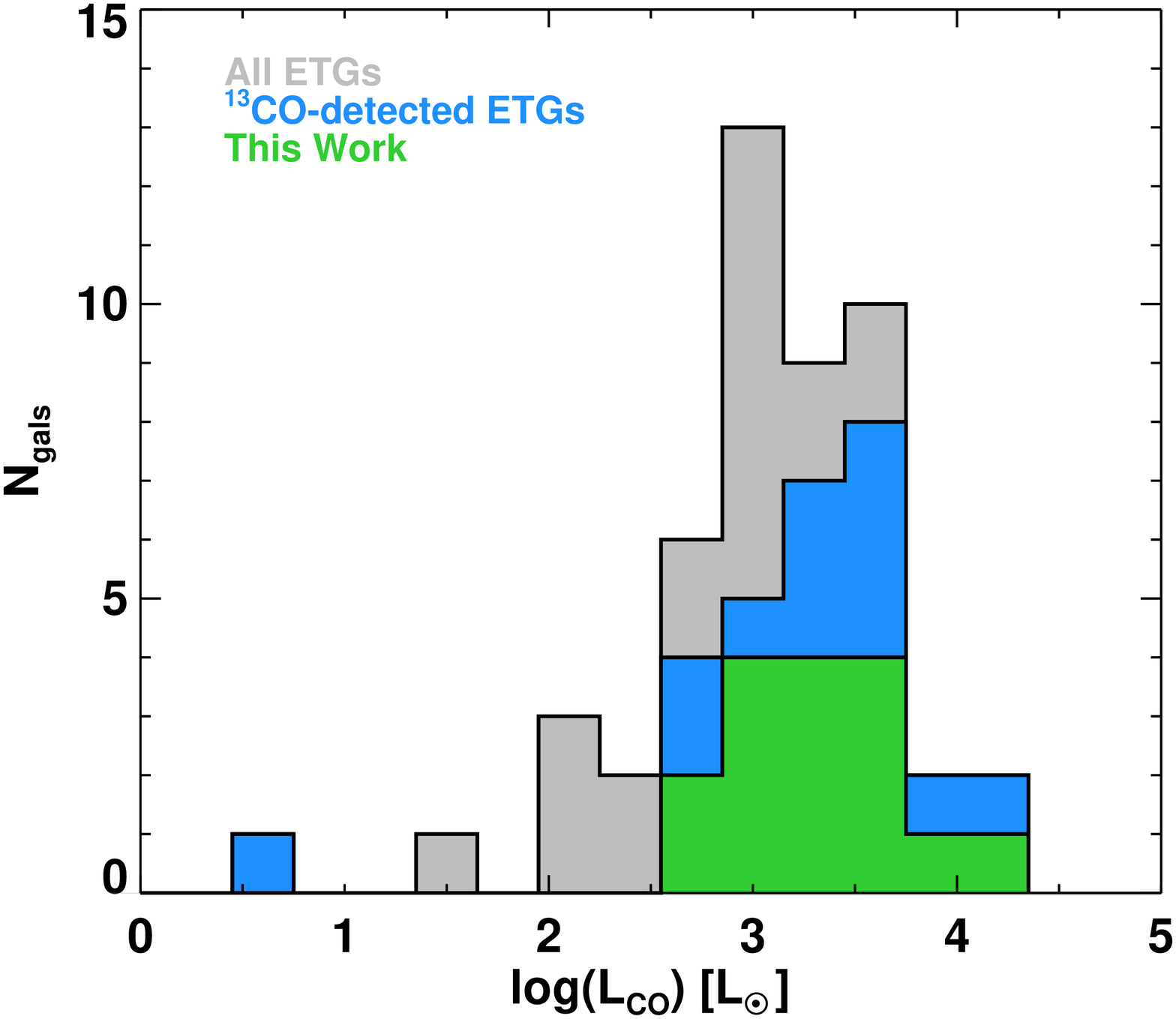}}
\subfigure{\includegraphics[width=0.48\textwidth,clip,trim=2.5cm 0.6cm 1.3cm 1.6cm]{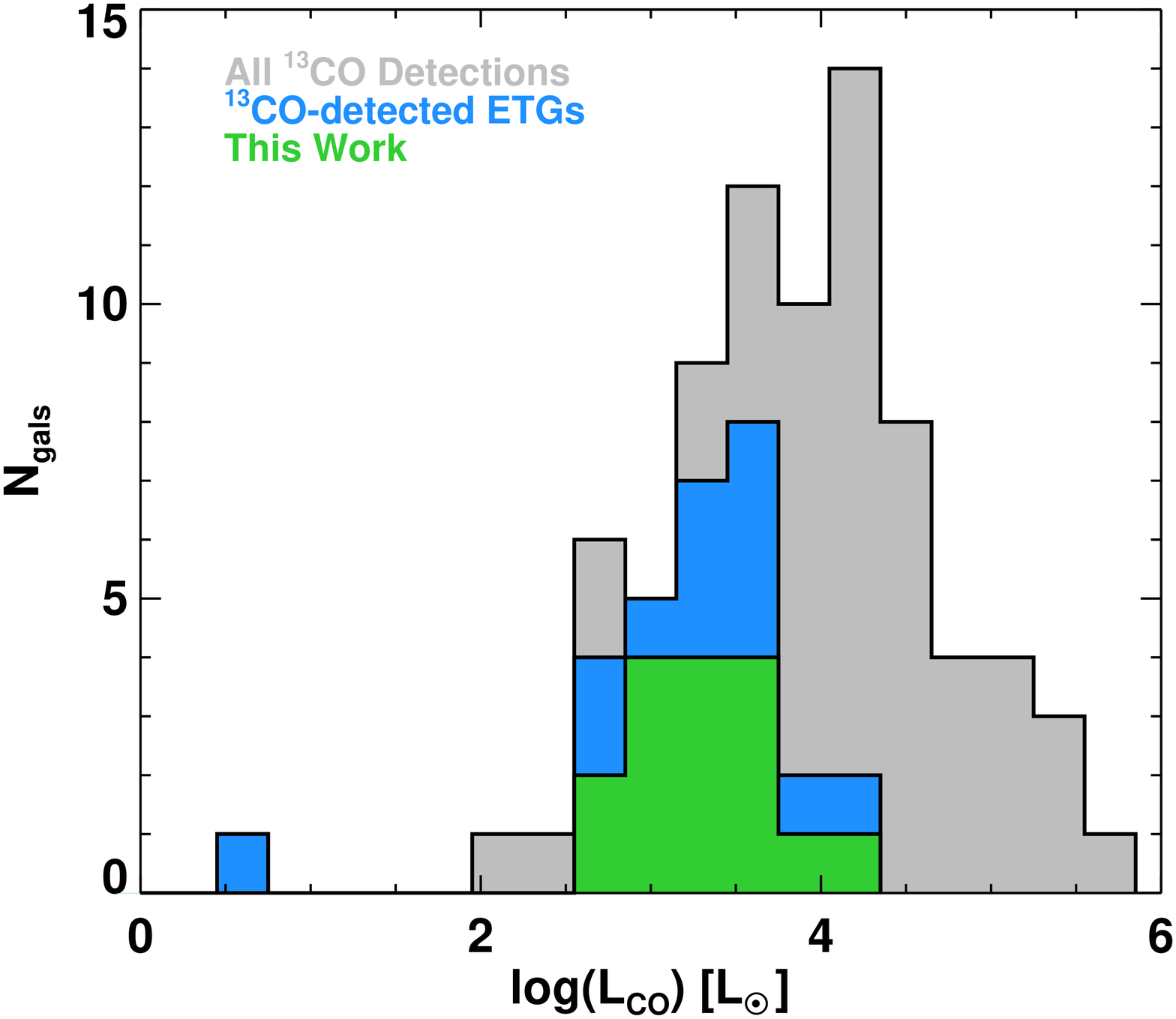}}
\caption{{\bf(Left):} Distribution of \twco\ luminosities (gray) among both \thco-detected ETGs (green) and this work (blue).  The \twco\ comparison sample is from the flux-limited \twco\ survey of \citet{young+11}.  The \thco\ sample represents a large percentage of ETGs detected at $L_{\rm CO} > 10^{2.5}$~L$_{\odot}$, though misses some of the faintest ETGs detected in \twco.  {\bf(Right):} Distribution of all galaxies with measured \rco, including LTGs (gray).  The \thco-detected ETGs are distinguished (blue) along with galaxies part of this work (green).  In this case, while the ETGs fall in the low $L_{\rm CO}$ end of the distribution, it appears that this is due to the fact that ETGs in general have lower $L_{\rm CO}$ (further detailed in \citealt{young+11}), rather than the \thco-detected ETGs being biased.  In both cases, the smallest value of $L_{\rm CO}$ is NGC\,404 from \citet{sage90}, at $L_{\rm CO} \approx 6$~L$_\odot$.}
\label{fig:comparison}
\end{figure*}

While early-type galaxies (ETGs) were originally assumed to be mainly devoid of molecular gas (e.g., \citealt{bower+92}), it has become increasingly apparent that these ``red and dead'' early-type galaxies often contain non-negligible amounts of cold gas \citep{wiklind+86,phillips+87,wiklind+89,sage+89,welch+sage03,sage+07}.  Large-scale searches for molecular gas in ETGs have indeed shown that a significant proportion of these galaxies do contain a molecular gas reservoirs (e.g., \citealt{combes+07, young+11}), though many of the early searches targeted relative small unrepresentative samples of ETGs, rather than unbiased, volume-limited samples.

The \atlas\ survey \citep{cappellari+11} provides a complete, unbiased sample of morphologically-selected ETGs in a large local volume.  All \atlas\ galaxies were searched for molecular gas by \citet{young+11}, with a measured detection rate of $22\pm3$\%.  The detection rate of molecular gas did not appear to be dependent on galaxy environment, with a similar fraction of group and cluster ETGs detected as field ETGs.  Follow-up efforts included CO imaging (e.g., \citealt{ybc08,crocker+11,alatalo+13}).  \citet{davis+11} found that despite the fact that Virgo galaxies are detected at a similar rate to field ETGs, the origin of the gas within Virgo galaxies appears to be exclusively internal, while a large fraction of field ETGs acquire their molecular gas via accretion of gas-rich companions.  

%%%%%%%%%%  Figure 2  %%%%%%%%%%
\begin{figure*}
\includegraphics[width=0.7\textwidth]{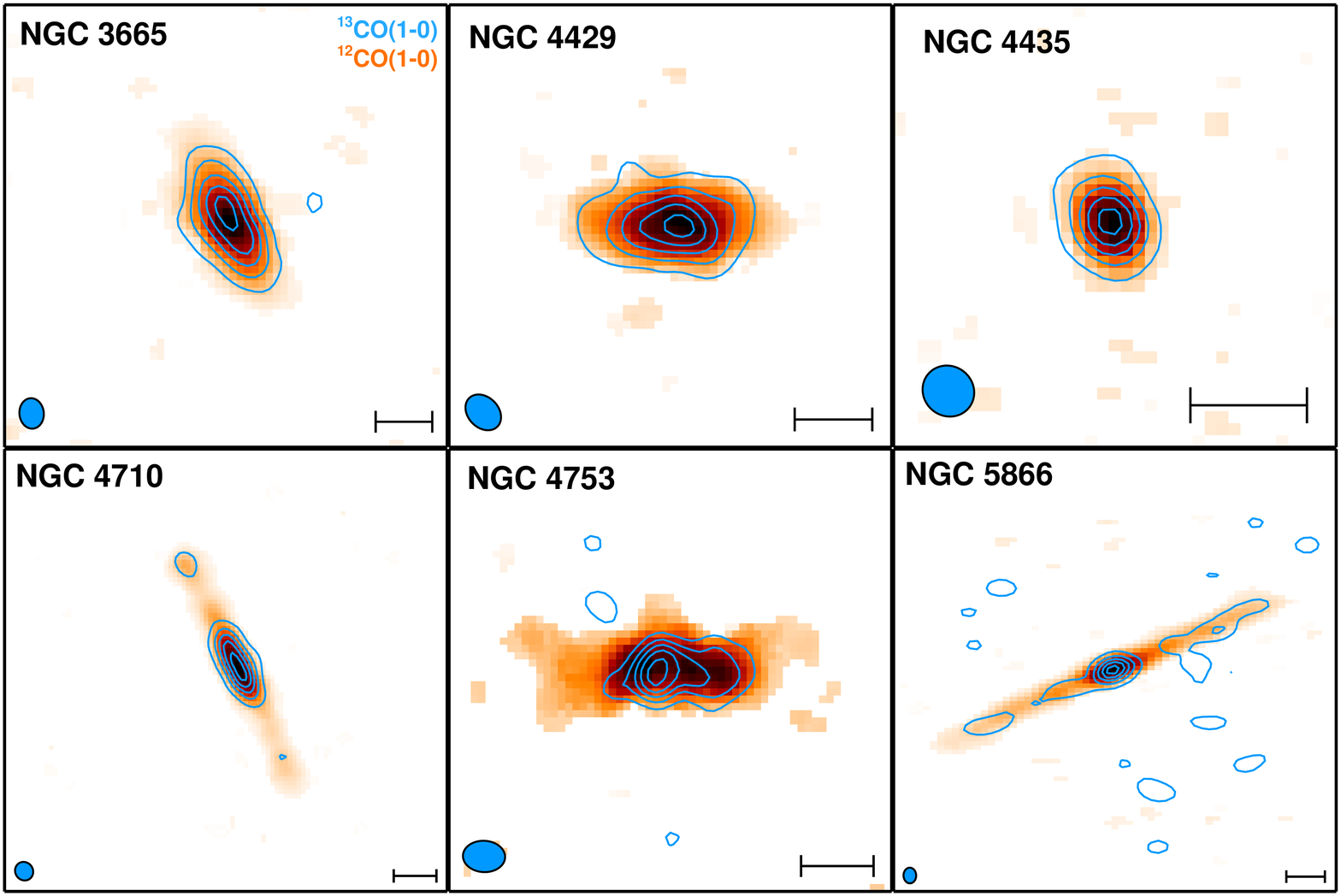}
\caption{Integrated intensity map of the \thco(1--0) (blue contours) overlaid upon the \twco\ moment0 map (colourscale) for NGC\,3665, NGC\,4429, NGC\,4435, NGC\,4710, NGC\,4753 and NGC\,5866, the galaxies with strong (see \S\ref{sec:images}) \thco\ detections.  The CARMA \thco(1--0) beam is shown at the bottom-left of each panel, indicating that the galaxies are well resolved, and a scale bar representing 10\arcsec\ is shown on the bottom-right.  Galaxies known to be in the Virgo cluster include NGC\,4429, NGC\,4435 and NGC\,4710 \citep{cappellari+11}.  NGC\,4753 is well-known to be in a group \citep{steiman-cameron+92} and NGC\,3665 and NGC\,5866 are within a galaxy association \citep{garcia93}.  The \thco\ levels are \hbox{[0.1, 0.3, 0.5, 0.7, 0.9]} times the maximum.}
\label{fig:co13_co}
\end{figure*}

\thco(1--0) has long been used as an optically thin tracer of \twco, and for the most part has been shown to be a faithful tracer of it \citep{young+sanders86,sage90,sage+91,aalto+95}.  The range of \thco/\twco\ ratios (hereafter \rco) seen in past studies (mainly in spirals, starbursts and mergers) was 0.06--0.4, with ongoing major mergers populating the lower end of the distribution \citep{aalto+91}.  \citet{sage90} studied the \thco(1--0) properties of 3 ETGs (NGC\,404, NGC\,4710 and NGC\,5195) with the National Radio Astronomy Observatory (NRAO) 12m single-dish, and concluded that this small sample of S0 galaxies seems to have a similar value of \rco\ as spiral galaxies on average.  NGC\,5195, known to have recently interacted with M51, has a much smaller \rco\ than the other ETGs, similar to what is seen in NGC\,1266 \citep{alatalo+11,crocker+12}, a peculiar ETG with compact, dense molecular gas \citep{a14_sfsupp} currently hosting an active galactic nucleus (AGN)-driven massive molecular outflow \citep{davis+12,a14_stelpop}.

%%%%%%%%%%  Table 1  %%%%%%%%%%
\begin{table*}
\caption{\thco(1--0) observations for the CARMA \atlas\ survey}
\begin{center}
\tabcolsep=0.16cm
\begin{tabular}{llccccccl}
\hline \hline
{\bf Name} & $\theta_{\rm maj}\times\theta_{\rm min}$ & Velocity Range & $\Delta v^\dagger$ & $F_{\rm^{13}CO}^\ddagger$ & $\frac{F_{\rm CARMA}}{F_{\rm 30m}}^\natural$& \rco\ &  $F_{\rm^{12}CO}^{\ddagger \flat}$ & $f_{60}/f_{100}$\\
& ~~~~~~~~~($''$) & (km s$^{-1}$) &  (km s$^{-1}$) & (Jy km s$^{-1}$) & & &(Jy km s$^{-1}$)\\
\hline
IC\,676 & $4.57\times4.00^\star$ & 1360 -- 1520 & 20 & $6.52\pm1.78$ & $0.96\pm0.3$ & $0.096\pm0.03$ & 67.9$\pm$5.8 &  $0.62^a$ \\
IC\,719 & $4.09\times3.74 $ & 1670 -- 2080 & 40 & $3.45\pm0.64$ & -- & $0.14\pm0.03$ & 23.8$\pm$2.9 &  $0.33$ \\
NGC\,2764 & $4.14\times3.79 $ & 2570 -- 2840 & 30 & $9.05\pm1.24$ & $1.4\pm0.2$ & $0.11\pm0.02$ & 79.7$\pm$4.2 &  $0.51^a$ \\
NGC\,3607 & $5.82\times5.19 $ & 680 -- 1210 & 40 & $8.57\pm0.69$ & $1.1\pm0.1$ & $0.16\pm0.02$ & 53.2$\pm$3.2 &  $0.79^a$ \\
NGC\,3619 & $4.65\times4.11 $ & 1280 -- 1770 & 80 & $< 1.25$ & -- & $< 0.083$ & 15.1$\pm$1.3 &  $0.21$ \\
NGC\,3626 & $4.14\times3.84 $ & 1320 -- 1670 & 50 & $3.89\pm0.83$ & -- & $0.097\pm0.02$ & 40.1$\pm$3.4 &  $0.56$$^b$ \\
NGC\,3665 & $4.42\times4.27 $ & 1750 -- 2400 & 50 & $20.0\pm0.80$ & $1.1\pm0.07$ & $0.22\pm0.01$ & 91.2$\pm$4.4 &  $0.30^a$ \\
NGC\,4324 & $4.93\times4.04 $ & 1520 -- 1770 & 40 & $3.06\pm0.50$ & -- & $0.15\pm0.03$ & 20.0$\pm$2.1 &  $0.21$ \\
NGC\,4429 & $5.12\times3.90 $ & 830 -- 1370 & 30 & $20.5\pm0.66$ & -- & $0.29\pm0.02$ & 70.4$\pm$3.2 &  $0.30$ \\
NGC\,4435 & $4.50\times4.19 $ & 560 -- 980 & 30 & $5.97\pm0.40$ & -- & $0.18\pm0.02$ & 32.6$\pm$2.6 &  $0.43$ \\
NGC\,4694 & $4.28\times3.39 $ & 1150 -- 1210 & 30 & $3.83\pm0.63$ & $2.3\pm0.7$ & $0.23\pm0.04$ & 17.0$\pm$0.8 &  $0.41^a$ \\
NGC\,4710 & $4.37\times4.07 $ & 940 -- 1310 & 30 & $51.5\pm2.47$ & $2.3\pm0.1$ & $0.15\pm0.007$ & 350.6$\pm$6.0 &  $0.39^a$ \\
NGC\,4753 & $5.81\times4.32 $ & 919 -- 1470 & 50 & $12.7\pm1.05$ & -- & $0.18\pm0.02$ & 70.2$\pm$4.9 &  $0.29$ \\
NGC\,5379 & $5.65\times3.90 $ & 1700 -- 1880 & 30 & $1.83\pm0.42$ & -- & $0.080\pm0.02$ & 22.9$\pm$2.3 &  $0.21$ \\
NGC\,5866 & $3.75\times3.27 $ & 500 -- 1010 & 30 & $34.1\pm2.59$ & $2.2\pm0.2$ & $0.14\pm0.01$ & 250.9$\pm$8.4 &  $0.27^a$ \\
PGC\,029321 & $4.10\times4.01 $ & 2760 -- 2910 & 30 & $< 1.46$ & -- & $< 0.074$ & 19.6$\pm$1.7 &  $1.3$ \\
PGC\,058114 & $4.72\times3.93 $ & 1370 -- 1640 & 30 & $5.68\pm1.13$ & $2.1\pm0.5$ & $0.077\pm0.02$ & 73.4$\pm$3.1 &  $0.78^a$ \\

\hline \hline
\end{tabular}
\end{center}
\label{tab:co13}
\raggedright {\footnotesize
$^\dagger$ Channel width of the \thco\ cube.\\
$^\ddagger$ Does not include the 20\% absolute millimeter flux calibration uncertainty.  Errors were calculated as discussed in \S\ref{sec:stacking}.\\
$^\natural$ 30m \thco\ integrated flux densities \citep{crocker+12}.\\
$^\flat$ \twco\ integrated flux densities from the CO cubes described in \citet{alatalo+13}, but derived with the same velocity resolution (and channels) as the \thco\ cubes to match the \thco\ resolution for our stacking analysis.\\
$^\star$The \twco\ map of IC\,676 is uniformly weighted in \citet{alatalo+13}, whereas the other \twco\ and \thco\ maps in this work and are naturally weighted\\
$^a f_{60}/f_{100}$ from \citet{crocker+12}.  Others values are calculated from IRAS fluxes.\\
$^b$IRAS data were unavailable, so the flux ratio was calculated based on the {\em Herschel} $f_{70}/f_{160}$ ratio (T. Bitsakis, private communication).}
\end{table*}

Dense gas (i.e., gas that is traced by HCN and HCO$^+$ emission) was observed in a much larger sample of ETGs through the SAURON \citep{sauron} and \atlas\ \citep{cappellari+11} surveys, that targeted the strongest \twco-detected \atlas\ galaxies with the Institut de Radioastronomie Millim\'etrique (IRAM) 30m single-dish telescope \citep{krips+10,crocker+12,davis+13b}.  In these works, there was some evidence that while ETGs show similar ratios of dense gas as spirals, the ratios may depend on environment, with Virgo galaxies showing boosted \rco\ and HCN/CO ratios compared to field galaxies.  With the \citet{crocker+12} sample, statistically significant correlations between \rco\ and absolute {\em K}-band luminosity, stellar population age, molecular to atomic gas ratio, dust morphology and dust temperature were found. The correlation between \rco\ and environment was not however statistically significant, perhaps due to the low numbers of galaxies in dense environments such as the Virgo cluster.
%At the time there were however very few Virgo ETGs observed in dense gas tracers, so \citet{crocker+12} were unable to confirm whether group and cluster environments were indeed causing a boost in \rco.

%As part of the Combined Array for Research in Millimeter Astronomy (CARMA) \atlas\ survey, designed to detect \twco\ \citep{alatalo+13}, we included simultaneous observations of \thco, to serve as a detection experiment for 17 of the ETGs as soon as CARMA instumentational capabilities allowed.  By stacking \thco\ data using the detected \twco\ signal as a 3D mask \citep{schruba+11}, we were able to detect \thco(1--0) in 15 of the 17 ETGs observed, and had strong enough detections of 5 of these galaxies to create images and do resolved studies.  Unlike the previous dense gas studies, the CARMA \thco\ survey probes much fainter CO fluxes, extending our understanding of dense molecular into more quiescent galaxies, as well as more galaxies in cluster and group environments.

Here we present the \thco(1--0) observations from the CARMA array.  In \S\ref{obs}, we discuss the simultaneous \twco(1--0) and \thco(1--0) observations, and the techniques that were used to reduce and analyse the \thco(1--0) data, including imaging strong detections and stacking all non-detections.  In \S\ref{sample}, we describe our sample galaxies and their properties.  In \S\ref{disc}, we discuss the implications of the \thco\ distributions and how our new detections inform upon what affects \rco\ in galaxies.  In \S\ref{conc}, we summarize our conclusions.  The \atlas\ sample properties are described in detail in \citet{cappellari+11}\footnote{http://www.purl.org/atlas3d}, and details specific to the CARMA observations are reported in \citet{alatalo+13}.  We use the cosmological parameters $H_0 = 70$~km~s$^{-1}$, $\Omega_M = 0.3$ and $\Lambda = 0.7$ \citep{wmap} throughout.

\section{The Sample}
\label{sample}
The CARMA sample has been extracted from the \atlas\ sample of galaxies, which is a volume-limited sample (D$<$42\,Mpc) of massive (M$_K$$<$-21.5\,mags) morphologically-selected early-type galaxies \citep{cappellari+11}.  The subsample of galaxies observed with CARMA have been selected among those detected in the \twco\ line by \citet{young+11}, who observed the complete \atlas\ sample.  CARMA then imaged all previously un-imaged CO-detected ETGs above a flux limit of 18.5\,Jy~km~s$^{-1}$ \citep{alatalo+13}. Figure~\ref{fig:comparison}a shows the distributions of \thco\ luminosities for \thco-detected ETGs from \citet{sage90,crocker+12} and this work compared to the unbiased, flux-limited survey of \citet{young+11}, indicating that \thco-detected ETGs tend to populate the high $L_{\rm CO}$ end of the distribution.  Given that CARMA observed the brightest \twco\ ETG detections from \citet{young+11}, the fact that \thco\ detections tend to populate the bright portion of the \twco\ distribution is unsurprising.  %For this work, galaxies observed with CARMA after the correlator upgrade contain simultaneous observations based on the hardware availability to simultaneously map \twco\ and \thco\ in separate sidebands.

Figure~\ref{fig:comparison} shows that when we draw comparisons to the available \thco-detected galaxies in the literature \citep{young+sanders86,sage+91,aalto+95}, that are either late-type galaxies (LTGs), starbursts or mergers, the \thco-detected ETGs fall to the lower end of the distribution of \twco\ luminosities, consistent with what is found in \twco-detected ETGs in general \citep{young+11}.  So our \thco\ ETG sample has properties entirely consistent with a population of \twco-bright ETGs.

It also appears clear that the galaxies added in this work do not fundamentally differ in their \twco\ properties from the galaxies considered in previous ETG work \citep{young+11}, and appear to be a robust addition to the \thco-detected ETGs.  If we look specifically at the representation of Virgo galaxies, we find that 7/26 (26$\pm$8\%) are Virgo members within the \thco-detected sample, compared to 12/56 (21$\pm$5\%) of the \twco-detected ETGs from \citet{young+11}.  Thus it does not appear that Virgo members are over- or under-represented in our sample.  \thco-detected ETGs were also selected from this subsample, using identical selection criteria to Virgo and thus are also neither over- nor under-represented in the sample.

Though the 7 Virgo members have comparable average properties to the non-Virgo \thco-detected ETGs (within the scatter), including stellar mass ($<M_{\rm K, Virgo}> = -23.6\pm0.9$ vs. $<M_{\rm K, non-Virgo}> = -22.9\pm1.1$) and molecular gas extent ($<R_{\rm CO}/R_{e,{\rm Virgo}}> = 0.45\pm0.45$ vs. $<R_{\rm CO}/R_{e,{\rm non-Virgo}}> = 0.67\pm0.43$; \citealt{davis+13b}), they do skew more massive with more compact molecular discs.  This is likely due to the small number of Virgo galaxies within this sample, but the scatter is large due to small number statistics.  Further \thco\ studies of more Virgo members should reduce the scatter.

\section{Observations and Results}
\label{obs}
\thco(1-0) observations were obtained for 17 of the 30 original CARMA \twco(1--0) galaxies.  The galaxies missing \thco(1-0) data are those imaged before Spring 2009, when upgrades to the CARMA receivers allowed for simultaneous imaging of both lines.  The total available bandwidth from CARMA in each line varies as well depending on the semester the source was observed, due to upgrades to the correlator.  Total bandwidths for each object are listed in Table 2 of \citet{alatalo+13}.  All data reduction and calibration were done simultaneously with the \twco\ imaging published in \citet{alatalo+13}.  Table \ref{tab:co13} presents the galaxies observed in \thco, along with \thco-specific observing parameters (including beam size, bandwidth and channel widths).  The calibrators used, and total observing time for each source, can be found in \citet{alatalo+13}. 

%%%%%%%%%%  Figure 3  %%%%%%%%%%
\begin{figure*}
\includegraphics[width=\textwidth]{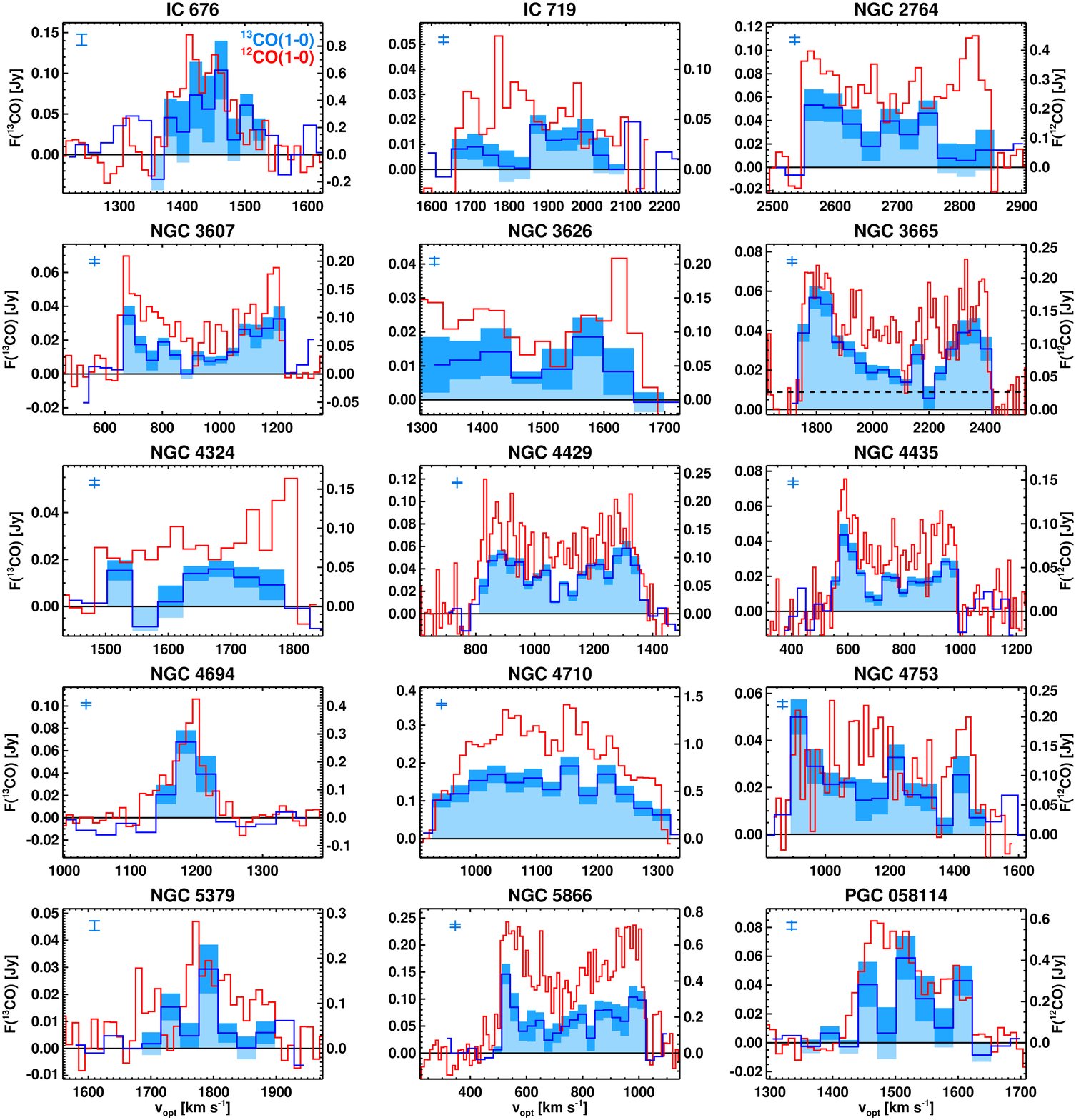}
\caption{Stacked \thco(1--0) spectra of the 15/17 detected galaxies in the CARMA \atlas\ \thco\ survey.  The dark blue shaded regions represent $\pm1\sigma$ uncertainties, using the noise calculation laid out in \S\ref{sec:stacking}, with the light blue representing the integrated signal within the 2D \twco(1--0) mask of each channel, compared with the \twco(1--0) spectra from \citet{alatalo+13} in red.  The un-shaded \thco\ spectra are those channels where the number of pixels in the \twco\ mask fall to fewer than total number of pixels of the beam.  In these channels, the spectrum is calculated by summing pixels in a contiguous region of the channel map the size of the beam area, and can be considered the baseline scatter for the stacked \thco\ spectra, with corresponding error bars in the top left of each panel.  The scale on the left of each panel is the \thco(1--0) line flux, while the scale on the right of each panel is the \twco(1--0) line flux.  The velocities are the optically-defined, local standard of rest ($v_{\rm opt, LSR}$) velocities of the molecular gas.   Total bandwidths from the CARMA observations are listed in Table 2 of \citet{alatalo+13}. In all cases except NGC\,3665, the continuum (reported in \citealt{alatalo+13}) is insubstantial compared to the stacked \thco\ spectra (and has been removed from all line data).  A dashed black line shows the continuum detection level of 8.98\,mJy in NGC\,3665.}
\label{fig:stacked}
\end{figure*}

\subsection{Strong \thco\ detections: \thco\ maps}
\label{sec:images}
The \thco(1--0) emission was strong enough (emission was detected with signal-to-noise ratio $>$3) to map 6 of the 17 observed galaxies. Moment maps were constructed based on the methods described in \citet{alatalo+13}, with a full velocity width determined by the \twco(1--0) line. Figure \ref{fig:co13_co} shows the six mapped galaxies: NGC\,3665, NGC\,4429, NGC\,4435, NGC\,4710, NGC\,4753 and NGC\,5866.  To test the agreement between the CARMA detections and the single dish detections from \citet{crocker+12}, we convolved the images of the three overlapping galaxies with strong \thco\ detections: NGC\,3665, NGC\,4710 and NGC\,5866, to the size of the IRAM 30m beam, and compared \rco, finding them to be in agreement within errors with \rco\ reported in \citet{crocker+12}.

For these six galaxies, we attempted to investigate whether \rco\ varies with position. We first regridded both the \thco\ and \twco\ maps so that the area of one square pixel equalled the area of the \thco\ beam, this using the Interactive Data Language (IDL) Astronomer's Library routine {\tt hcongrid}\footnote{http://idlastro.gsfc.nasa.gov/ftp/pro/astrom/hcongrid.pro}, with bilinear interpolation of all points.  We then measured both a nuclear (from the regridded pixel corresponding to the 2 Micron All Sky Survey K$_{\rm s}$ band-determined nuclear position; \citealt{2mass}) and a non-nuclear \rco\ value (from the remaining \thco-detected pixels).  To evaluate the uncertainty on the \rco\ measurements, we first evaluated the uncertainty on each integrated flux by measuring the root mean square noise (RMS) in the regridded \twco\ and \thco\ integrated intensity maps, respectively, taking the standard deviation of the pixels outside of the region where \twco\ is detected, and multiplying by $\sqrt{N}$, where $N$ is the number of pixels in each region (one for nuclear regions; the total number of detected off-nuclear pixels for the non-nuclear regions).  These \rco\ gradient measurements are listed in Table~\ref{tab:resolved_co13} and discussed in Section~\ref{disc}.

%%%%%%%%%%  Table 2  %%%%%%%%%%
\begin{table}
\caption{Resolved \rco\ values}
\begin{tabular}{lcc}
\hline \hline
{\bf Name} & \rco$_{\rm ,nuc}$ & \rco$_{\rm ,ext}$\\
\hline
NGC\,3665 & $0.24\pm0.01$ & $0.29\pm0.01$ \\
NGC\,4429 & $0.29\pm0.01$ & $0.41\pm0.02$ \\
NGC\,4435 & $0.35\pm0.01$ & $0.34\pm0.01$ \\
NGC\,4710 & $0.18\pm0.01$ & $0.17\pm0.01$ \\
NGC\,4753 & $0.15\pm0.03$ & $0.10\pm0.03$ \\
NGC\,5866 & $0.06\pm0.02$ & $0.08\pm0.02$ \\
\hline \hline
\end{tabular}
\label{tab:resolved_co13}
\raggedright
\end{table}

\subsection{Weak \thco\ detection: \thco\ spectra}
\label{sec:stacking}
To extract \thco\ integrated flux densities for the fainter galaxies, where \thco\ is not detected in individual beams, we utilize our prior knowledge of the molecular gas distribution from the brighter \twco\ emission.  This technique is also used to extract integrated flux densities for the strong \thco\ detections. We used a smoothed \twco\ datacube (created to correspond both to the weighting scheme as well as velocity properties of the \thco\ maps) for each galaxy as an priori dataset, and for each channel determined the two-dimensional region over which the \twco\ line is detected with a signal-to-noise ratio above 1.5.  This 3D region then represents a \thco\ detection mask (see \citealt{schruba+11} for details of a similar method). For each galaxy, we then sum the \thco\ datacube over its detection mask channel by channel, to obtain an integrated spectrum (shown in Fig. \ref{fig:stacked}) and thereafter an line flux\footnote{Line flux is defined as being the emission detected by the interferometer in the \twco\ masking aperture, summed over the channel range determined by the observer}. (tabulated in Table\,\ref{tab:co13}) by summing the channels with emission (shaded blue in Figure~\ref{fig:stacked}).  The un-shaded portion of the \thco\ spectra in Fig.~\ref{fig:stacked} are channels where the number of pixels in the \twco\ mask fall to fewer than the number of pixels in the beam.  In these channels, the channel flux is calculated by summing pixels in a contiguous region of the channel map the size of the beam area, and can be considered the baseline scatter for the stacked  \thco\ spectra.

The RMS per channel was determined by calculating the standard deviation of the line flux in an area equivalent to the mask, created by the {\sc miriad} task {\tt restor} in ``residual'' mode, and multiplying by $\sqrt{N}$, where $N$ is the total size of the CO mask area normalized to the beam area.  The RMS on the total line flux was then calculated by multiplying the noise per channel by the width of the channel $\Delta \nu$ (shown in Table \ref{tab:co13}) and multiplying by the square root of the sum of the noise squared in each individual channel (the shaded channels in Fig. \ref{fig:stacked}):

\begin{equation*}
\sigma(F_{\rm ^{13}CO}) = \Delta v \left(\sum\limits_{i = 1}^{\rm N} \sigma^2_{i,{\rm ^{13}CO}}\right)^{1/2},
\end{equation*}

%\[ \sigma(F_{\rm ^{13}CO}) = \Delta v \left(\sum\limits_{i = 1}^{\rm N} \sigma^2_{i,{\rm ^{13}CO}}\right)^{1/2} \]

\noindent where $i$ is the channel number.

There is a further 20\% absolute flux calibration uncertainty associated with millimeter observations, due to the time variability of millimeter flux calibration sources.  While this 20\% uncertainty impacts the total \thco\ line flux measurements, it does not factor into \rco, given that all \twco\ and \thco\ observations were taken simultaneously.  A further 30\% up-correction was made to the RMS to attempt to account for the oversampling of our interferometric maps.  The 30\% correction was derived by calculating the noise in the oversampled (1$''$ pixel) maps, and comparing that noise to the RMS derived from maps whose pixel area equals that of the synthesized beam.  The RMS per channel for each galaxy  is shown in the upper-left corner of each panel.  %A more detailed analysis of oversampled noise relations in interferometric data can be found in Topal et al. (2014), in preparation.  This noise per channel was then integrated over the velocity width of the detection (listed in Table \ref{tab:co13}), to obtain the total rms noise across the spectrum.  

%%%%%%%%%%  Table 3 %%%%%%%%%%
\begin{table}
\caption{Average \rco\ properties of each subsample of galaxies}
\begin{center}
\begin{tabular}{rccc}
\hline \hline
{\bf \thco\ Type} & {\textbf{N$_{\rm gals}$}} & {\bf Mean} & {\bf Std. Deviation}\\
\hline
{\bf All \thco} & 72 & 0.111 & 0.046\\
{\bf All Field}$^\dagger$ & 62 & 0.096 & 0.034\\
{\bf Spirals} & 32 & 0.098 & 0.029\\
{\bf Mergers} & 10 & 0.086 & 0.020\\
{\bf Field galaxies ($f_{60}/f_{100} < 0.5)$} & 32 & 0.112 & 0.036\\
\hline
{\bf All ETGs} & 27 & 0.128 & 0.081\\
\hline
{\bf Interacting ETGs} & 8 & 0.068 & 0.026\\
{\bf Settled ETGs} & 19 & 0.154 & 0.074\\
\hline
{\bf Virgo ETGs} & 7 & 0.221 & 0.064\\
{\bf Field ETGs} & 20 & 0.096 & 0.052\\
{\bf Field ETGs ($f_{60}/f_{100} < 0.5)$} & 8 & 0.120 & 0.066\\
\hline \hline
\end{tabular}
\end{center}
\label{tab:avg_co13}
\raggedright {\footnotesize
$^\dagger$The field sample includes all galaxies not in the Virgo cluster\\}
\end{table}

Table \ref{tab:co13} lists the galaxies that were observed in \thco\ during the CARMA \atlas\ survey, with the derived integrated flux densities and \rco\ values from the stacked spectra.  Of the 17 galaxies, 15 are detected using the a priori masking technique.  Figure \ref{fig:stacked} presents the stacked \thco\ spectra of the 15 detected galaxies.  Comparing the line shapes of \thco\ (blue) and \twco\ (red), we see that they are generally in good agreement, supporting our claim that the stacking technique as well as our chosen mask threshold have successfully detected \thco\ emission.  

%\thco\ spectra are obtained by summing over the \twco\ detection mask channel by channel. These are presented in Fig.~\ref{fig:all_co13}.  %The fact that the \thco\ has similar line profiles to the \twco\ in most of the detected galaxies in unsurprising, given our use of the apriori \twco\ channel maps as masks.   The fact that most of the stacked spectra show that the line profile exceeds the expected statistical noise seems to confirm that we have recovered \thco\ emission from the stacking method.

\citet{crocker+12} published IRAM 30m single-dish spectra of \thco(1--0) and \thco(2--1) for 18 of the \atlas\ galaxies.  Our sample overlaps the \citet{crocker+12} sample in the case of 8 galaxies: IC\,676, NGC\,2764, NGC\,3607, NGC\,3665, NGC\,4694, NGC\,4710, NGC\,5866 and PGC\,058114.  We detect all overlapping galaxies at a signal-to-noise ratio $>3$.  CARMA recovers more emission than the IRAM 30m in all cases except IC0676, equal within the uncertainty (see Table \ref{tab:co13}).  In many cases, this is likely because the \thco\ emission subtends a larger angular size than the 30m beam (as was often the case for the \twco\ emission published in \citealt{alatalo+13}).  Because \thco\ and \twco\ were measured simultaneously with CARMA, \rco\ does not suffer from flux calibration uncertainties.    We therefore elect to use the \thco\ integrated flux densities from the current work for all overlapping galaxies.

Radio continuum pollution leading to enhanced \thco\ fluxes is not likely to be an issue.  NGC\,3665 and NGC\,4710 are the only two galaxies in this sample with detected continuum, at $\approx 9$ and 4\,mJy, respectively, at 110\,GHz.  The \thco\ data processing includes a continuum subtraction step (detailed in \citealt{alatalo+13}).
%\thco, which is 3 times larger than the detected flux density per channel seen in Fig. \ref{fig:stacked} (this level is seen as an orange line).  Given the faintness of \thco, it is possible that an overestimation of the continuum, and subsequent subtraction, could remove a substantial fraction of the faint emission. 
%As the interferometer is also able to pick up \thco\ emission from an area larger than the $23''$ IRAM 30m beam

Table \ref{tab:resolved_co13} compares \rco\ in the nuclear and extended regions of NGC\,3665, 4429, 4435, 4710, 4753 and 5866 (see Section \ref{sec:images}).  In the first four, \rco\ is larger in both the nuclear and extended region than in the stacked spectrum, and vice versa for NGC\,4753 and NGC\,5866.  The discrepancies between stacking and resolved \rco\ measurements can be explained by many effects.  Stacking is able to detect fainter \thco\ emission from a larger area than the resolved data.  Compared to well-resolved objects, where the brightest \thco\ peaks were able to fill the beam, the average \rco\ from stacking would be lower than from the resolved maps.  On the other hand, in cases where CARMA has marginally detected \thco\ in the maps, stacking would be able to increase the total \thco\ flux detected, resulting in a higher \rco\ for stacking than mapping.  Deeper \thco\ observations might be able to fully explain this discrepancy, but are beyond the scope of this paper.

\section{Discussion}
\label{disc}
\subsection{Distribution of \thco\ in the galaxies}
Figure \ref{fig:co13_co} shows the spatial distribution of the \thco\ emission, compared to the \twco\, in \atlas\ galaxies.  Only the galaxies with the brightest \thco\ are detectable in a spatially-resolved way.  There are many features worthy of notice in the \thco\ distributions.  

NGC\,4710 and NGC\,5866 have strong, edge-on bars whose kinematic features were detected in the \twco\ emission (see \S4.2.2 in \citealt{alatalo+13}).  In these cases, the \thco\ preferentially traces the bar region, particularly a nuclear disc or ring and an inner ring, the locations of the largest concentrations of molecular gas in barred systems \citep{ath+bureau99}.  NGC\,4710 and NGC\,5866 were also among the handful of galaxies detected in OH absorption by \citet{mcbride+14}, confirming high optical depths.  The \thco\ detections at the disc edges further confirm the conclusions of \citet{alatalo+13} and \citet{davis+13a}, that these systems are bar+ring systems, as the \thco\ there is likely tracing the portions of the edge-on rings with higher column densities (edge brightening).  Topal et al. (in preparation) confirm this and study the spatially-resolved bar signatures using a multitude of high-density molecular tracers in both galaxies.
%It is also of note that the five brightest detections of \thco\ are in galaxies that are in groups, or clusters.  NGC\,4429, NGC\,4435, and NGC\,4710 are Virgo galaxies, NGC\,4753 is known to be the dominant galaxy within a forming group \citep{steiman-cameron+92} and NGC\,5866 is part of an association \citep{garcia93}.
% If these galaxies were both bar+discs, one would not expect such an enhancement on the outermost edge, and instead likely would expect the opposite to be true, with limb-decrements due to peering through smaller columns of gas.  

The Virgo galaxies NGC\,4429 and NGC\,4435 are robustly detected in \thco\ throughout the extent of the \twco\ detections.  In both cases, \thco\ faithfullly traces \twco.  NGC\,4753, known to be at the centre of a developing group \citep{steiman-cameron+92}, also seems to have \thco\ that traces the brightest portions of \twco, though it is likely that beyond these points the \thco\ emission simply drops below the detection threshold (rather than being absent).

%%%%%%%%%%  Figure 4  %%%%%%%%%%
\begin{figure}
\includegraphics[width=0.48\textwidth,clip,trim=1.1cm 1cm 0.9cm 1cm]{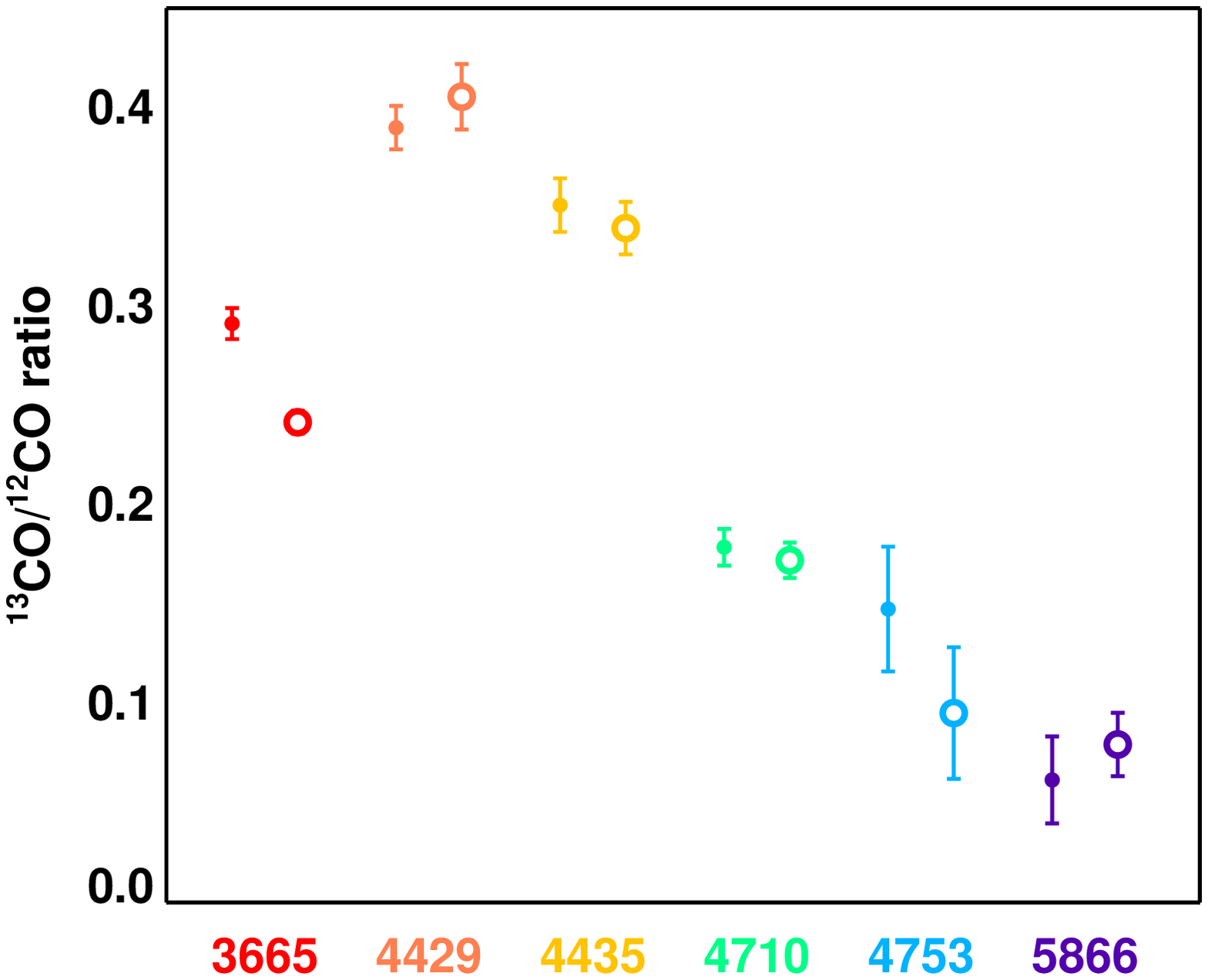}
\caption{Resolved \rco\ in the six strong \thco\ detections amongst the \atlas\ galaxies, separating the nucleus (points) and the surrounding gas (open circle). These numbers are likely upper limits for integrated \rco, as only \thco\ integrated flux densities exceeding our detection threshold are represented.}
\label{fig:resolved_co13}
\end{figure}

\subsection{Radial gradients in \rco}
Figure \ref{fig:resolved_co13} show \rco\ in nuclear and surrounding gas in the 6 galaxies mapped, with the points representing the nuclear pixels and the open circles representing the extended emission (with resolved \rco\ values shown in Table \ref{tab:resolved_co13}).  As discussed in \S\ref{sec:stacking}, ratios measured from the images differ slightly from those measured with the stacked spectra, likely due to the images missing faint \thco\ emission picked up in the stacked spectra.  Within the selected pixels, these ratios are thus likely reasonable, but they are not necessarily representative of the entire galaxy, where the stacking method is better able to recover emission below the local detection threshold.

In spiral galaxies, \citet{paglione+01} found a \rco\ gradient, increasing from the nuclear region to the outlying regions on average, though this result had large scatter.  Molecular discs in ETGs are much less extended than those in spirals \citep{davis+13b} in absolute terms, so it is possible that a \rco\ variation will be much less pronounced in ETG discs, either due to resolution or because the gas is better mixed (i.e. shorter dynamical times).  In the 6 spatially-resolved CARMA detections (shown in Fig. \ref{fig:resolved_co13}), there is no decipherable trend, and in 5 of the 6 objects the nuclear and non-nuclear \rco\ values are consistent within the errors.  NGC\,3665 is the single object in which the trend is larger than the uncertainties, but follows the opposite trend from spirals, and could be due to sensitivity limitations in detecting fainter \thco\ flux. There might be a slight increase in \rco\ from the nuclear region to the off-nuclear regions in NGC\,4429 and NGC\,5866, but \rco\ remains flat in NGC\,4710 and decreases in NGC\,3665, NGC\,4435 and NGC\,4753.  It is however possible that we are missing a non-negligible amount of faint extended emission, due to the sensitivity limitations of the CARMA sample.  As mentioned above (\S\ref{sec:stacking}), the disagreement in \rco\ between the image measurements and the stacked measurements is evidence that this might be the case.

We therefore do not see any evidence for an increase in \rco\ across galactic discs in this subset of ETGs, in contrast to what has been seen in spirals, but we cannot rule out that such a gradient could exist, given the sensitivity as well as resolution limitations of our observations. Deeper, higher resolution follow-up observations should more robustly establish whether ETGs indeed lack a \rco\ gradient, whether this is an environmental effect, or whether our result is due to sensitivity issues.

\subsection{Variations of \rco\ based on galaxy type and environment}
\label{sec:rco_var}
\begin{figure*}
\subfigure{\includegraphics[width=\textwidth]{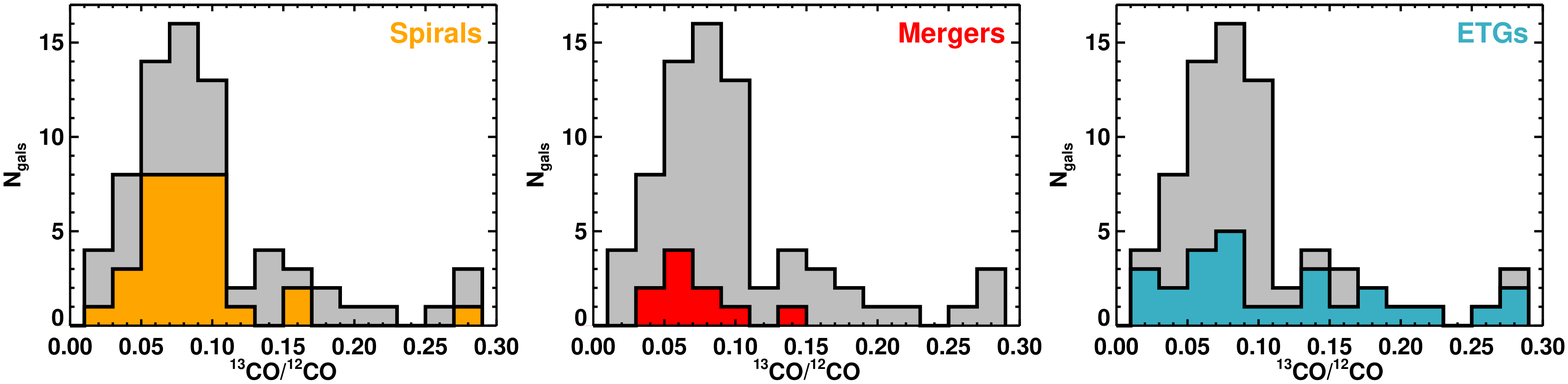}}
\subfigure{\includegraphics[width=\textwidth]{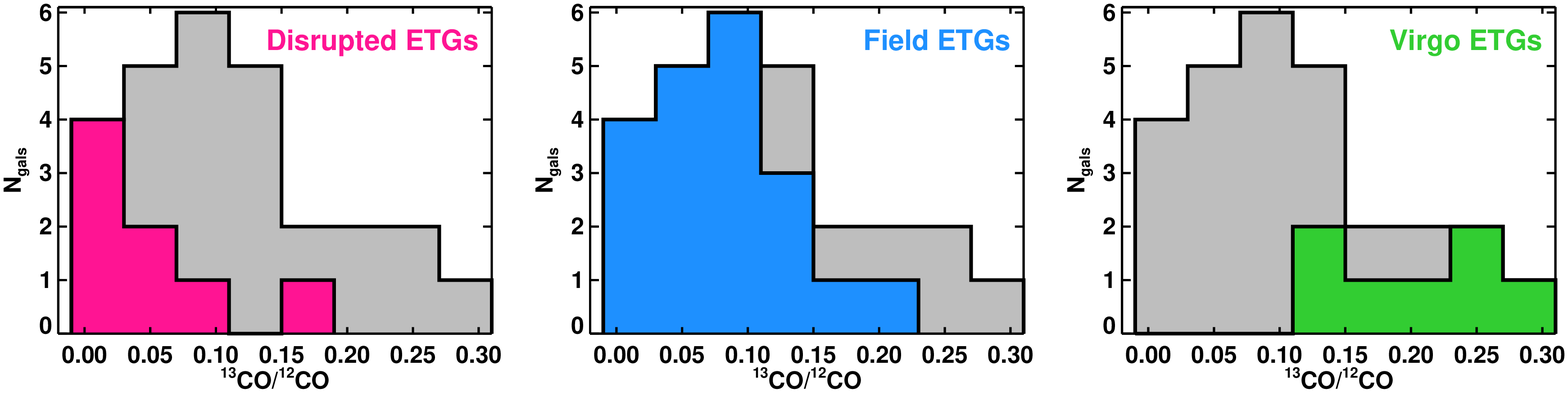}}
\caption{{\bf(Top):} Distribution of all galaxies with \rco (gray), compared with the distribution of spirals (orange), mergers (red) and ETGs (turquoise) only.  Both spirals and mergers seem to have an average \rco$\lesssim0.1$, whereas ETGs have a much flatter distribution.  {\bf(Bottom):} Distribution of all ETGs with \rco (gray), compared to the distribution of CO-disrupted ETGs (pink), field ETGs (blue), and Virgo ETGs (green) only.  Disrupted ETGs tend to have the smallest \rco\ values, while field ETGs are consistent with spirals.  Virgo ETGs account for the majority of large \rco.}
\label{fig:histograms}
\end{figure*}

Figure \ref{fig:histograms} presents the distributions of \rco\ as a function of galaxy types, in the new CARMA detections of Virgo and non-Virgo ETGs, the \citet{crocker+12} ETGs, NGC\,404 and NGC\,5195 from \citet{sage90}, as well as literature spirals and mergers \citep{young+sanders86,sage+91,aalto+95}. Figure \ref{fig:histograms} also shows the \rco\ distribution as a function of environment in the case of ETGs only (as there is no Virgo spiral with \thco\ data currently in the literature).  Spirals and mergers tend to have \rco$<0.1$, while ETGs are distributed much more evenly as a function of \rco.

We use the morphological classifications from \citet{alatalo+13} to differentiate between settled and disturbed gas morphologies, though NGC\,1266 was re-classified as disturbed, since the turbulent nature of the molecular gas is visible in the \twco\ and CS(2--1) lines using higher resolution imaging \citep{alatalo+11,a14_sfsupp}.  The \twco\ in NGC\,404 was classified as a ring \citep{wiklind+90}, and is therefore settled, while that in NGC\,5195 would be classified as mildly disrupted \citep{kohno+02}, and is considered disturbed.  The \rco\ value within ETGs seems to have a strong dependence on the disruption of the molecular gas, with disrupted gas distributions tending toward the lowest \rco\.  When ETGs are split by environment, the Virgo ETGs represent the majority of ETGs with \rco$>0.1$, while field ETGs average less than \rco=0.1.  The distributions not only show that environment seems to play a role, but also that disrupted molecular gas distributions tend toward lower \rco\ in ETGs.

%%%%%%%%%%  Figure 5  %%%%%%%%%%

\begin{figure*}
\includegraphics[width=0.8\textwidth]{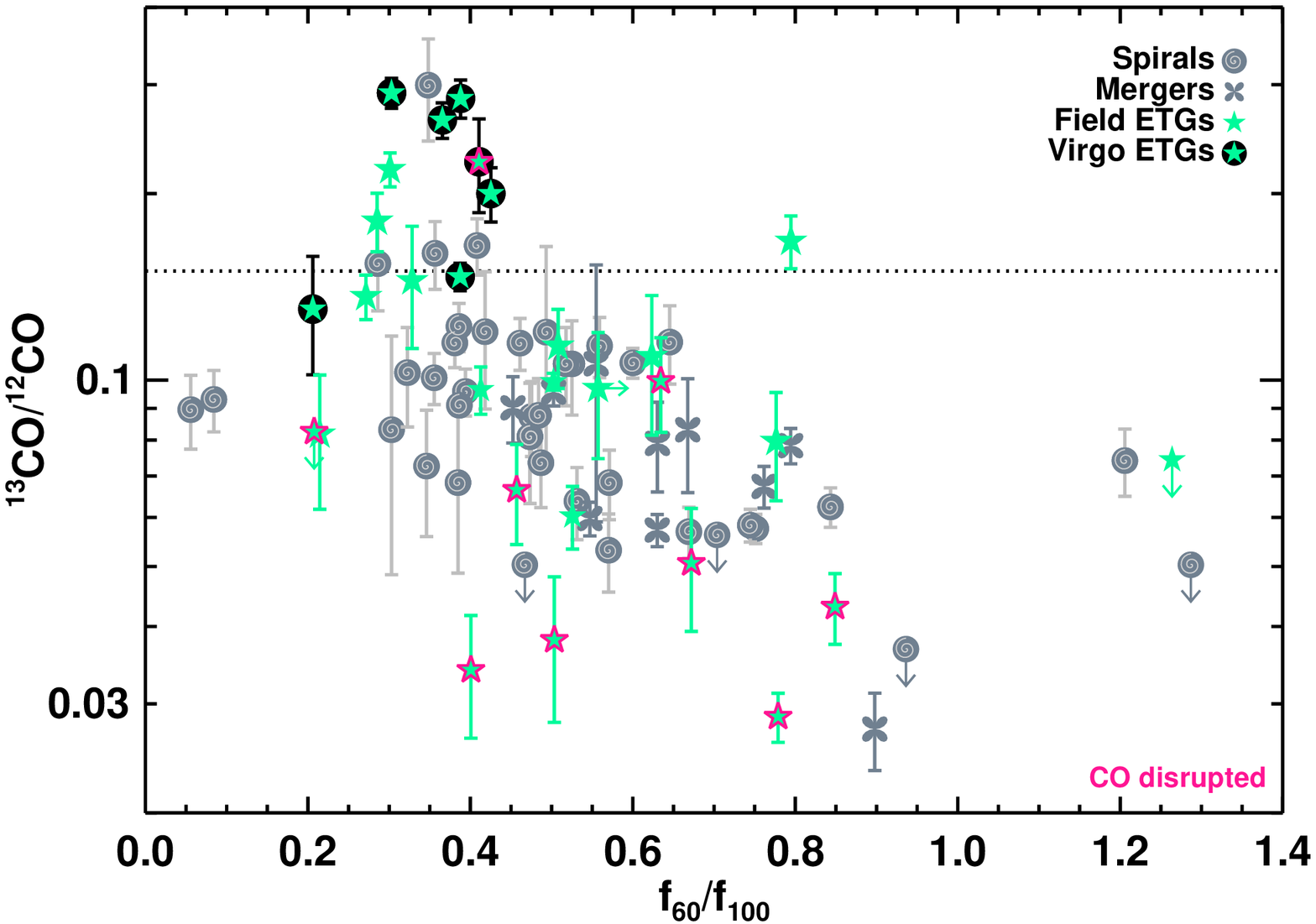}
\caption{Integrated $^{13}$CO/$^{12}$CO ratio for ETGs (green stars) are shown as a function of the 60--to--100$\mu$m flux ratio ($f_{60}/f_{100}$), and compared to spiral galaxies (spirals), and mergers (clovers).  Upper limits are shown as downward facing arrows. Literature spiral galaxies and mergers are from \citet{young+sanders86}, \citet{sage+91}, and \citet{aalto+95}, with ETGs (green) from \citet{sage90}, \citealt{crocker+12}, and this work.  The two ETGs from \citet{sage90} are NGC\,5195 (the minor interaction companion to M\,51; \citealt{kohno+02}) and NGC\,404.  The black dotted horizontal line shows the location of \rco=0.15.  Virgo ETGs are represented by green stars with a circular black background. All spirals are field spirals, as there is no Virgo spiral with \rco\ literature data.  While a few ETGs from \citet{crocker+12} have elevated \rco\ ratios, the majority have ratios similar to those of spirals and mergers from previous work.  Moreover, these ETGs also follow the trend with dust temperature $f_{60}/f_{100}$, with the highest \rco\ corresponding to the coldest dust temperatures.  ETGs (stars) with a magenta outline represent those that are morphologically classified as either mildly or strongly disrupted in \citet{alatalo+13}.  It is clear that ETGs with this classification are much more likely to have a low \rco, except in the case of NGC\,4694, which is also a Virgo galaxy. While two of the high dust temperature and low \rco\ ETGs from our study contain an AGN (NGC\,3607 and PGC\,029321), PGC\,058114, which also has a high dust temperature (but also a high \rco), does not appear to (Nyland et al., in preparation).}
\label{fig:all_co13}
\end{figure*}

Figure \ref{fig:all_co13} compares \rco\ with the dust temperature, represented by the 60-to-100 $\mu$m flux ratio.  \citet{young+sanders86} discussed the dependence of \rco\ on the dust temperature, with lower values of \rco\ corresponding to higher values of $f_{60}/f_{100}$, further confirmed by the observations of \citet{sage+91} who increased the sample size.  \citet{aalto+95} drastically increased the number of spirals for which \thco\ observations exist, confirming the trend as well as hypothesizing that turbulence within the molecular gas might influence \rco.  Warmer dust temperatures are a reasonable tracer of turbulence in the underlying molecular gas, as an abundance of young stars will provide both UV photons to heat the dust and momentum to stir up the molecular gas. \citep{ogle+10}.  However, the presence of Virgo galaxies at low $f_{60}/f_{100}$ also having the highest \rco\ indicate that environment could also be a factor in setting \rco.

\citet{crocker+12} found evidence that Virgo galaxies and those galaxies in groups have boosted \rco\ values, but due to small number statistics were not able to confirm that result.  The CARMA \thco\ detections add many more ETGs to the \citet{crocker+12} work, including an additional 3 Virgo galaxies.  The addition of these ETGs not only strengthens the trend, but appears to also further support the suggestion that turbulence might influence \rco\ \citep{young+sanders86,aalto+95}.  Magenta outlines in Fig. \ref{fig:all_co13} correspond to ETGs with disturbed molecular gas. The disturbed class of ETGs almost exclusively appears at values of \rco$<0.1$, the only exception being NGC\,4694 that also is a member of the Virgo cluster.

In spirals and mergers, there appears to be a ceiling in \rco\ of around 0.15 for the majority of sources.  Once the ETGs are included, a large proportion of them inhabit the \rco$>0.15$ region, with Virgo ETGs exclusively lying above this threshold. NGC\,4710 (identified as a Virgo member in \citealt{cappellari+11}) is found to have the lowest \rco\ in the Virgo ETG sample, of 0.15, though it is of note that while NGC\,4710 has a velocity similar to other Virgo cluster members, it is not formally within the virial radius of Virgo.  \rco\ in the CARMA sample is on average larger than that in all other samples discussed, due in part to the fact that the CARMA observations include more ETGs with predominantly settled gas distributions (which were not among the brightest \twco\ detections in \atlas\ first followed up in \thco\ with the IRAM 30m; \citealt{crocker+12}).  This inclusion of the ETGs with settled molecular gas seems have increased the population of objects in the \rco$>0.15$ region by over a factor of two.

Table \ref{tab:avg_co13} gives the \rco\ mean and standard deviation of each subsample of galaxies shown in Fig. \ref{fig:histograms}, including isolated galaxies based on environment, morphology and dust temperature.  Overall, Virgo ETGs have a larger mean \rco\ value than all other subsamples and are distinct (means inconsistent within the standard deviations) from all other subsamples except for cold dust field ETGs (of which there are only 8).  While these two populations agree within the errors, the cold dust field ETG population is much more similar to the complete cold dust field sample than to Virgo ETGs.  Thus, while we cannot completely rule out that the variation in \rco\ that we observe is solely dependent on molecular gas turbulence (as traced by dust temperature), it does appear that \rco\ in Virgo is boosted.

%%%%%%%%%%  Figure 6 %%%%%%%%%%

\begin{figure*}
\raggedright
\subfigure{\includegraphics[width=\textwidth]{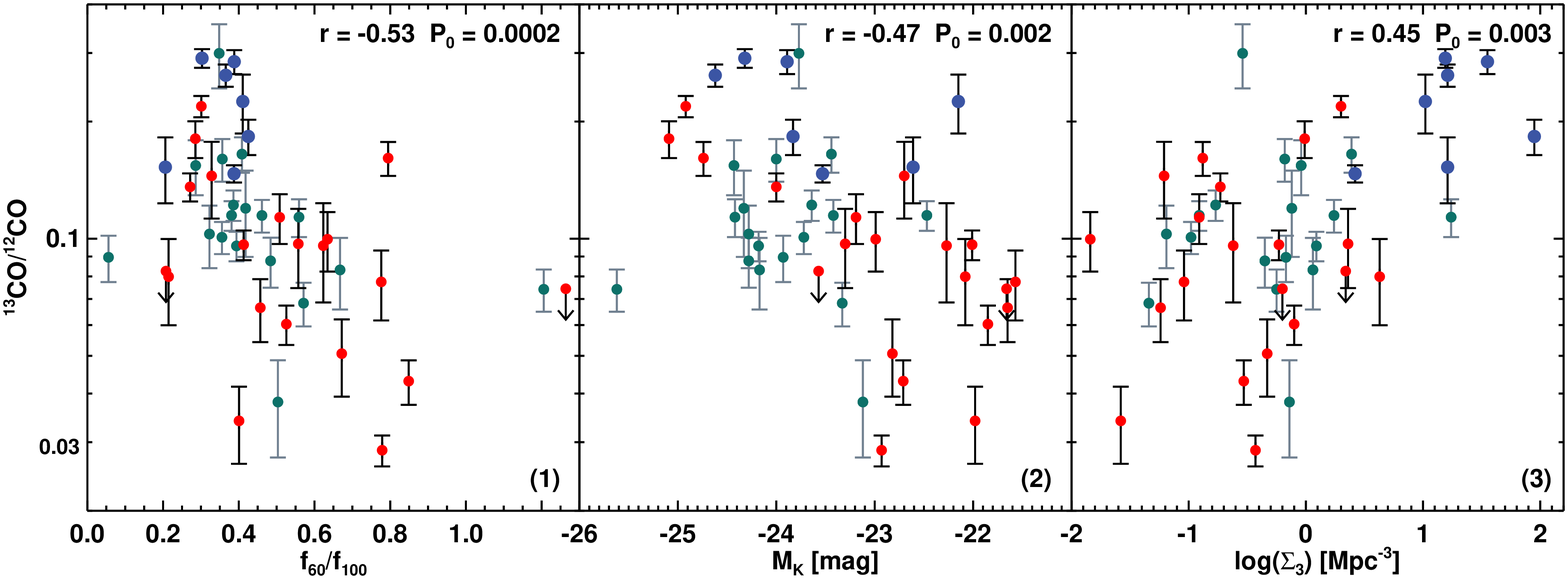}}
\subfigure{\includegraphics[height=7cm]{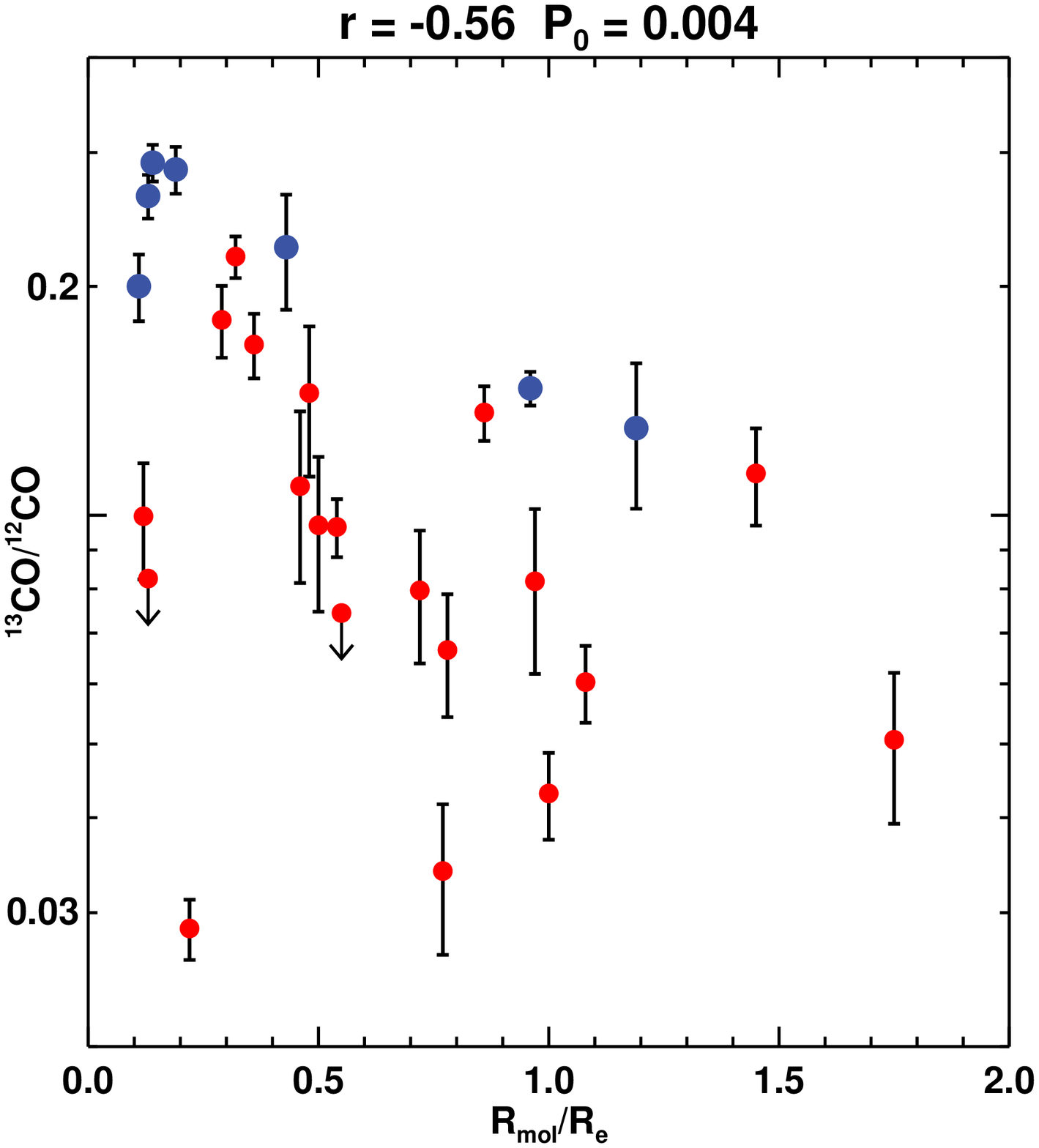}}
\caption{\rco\ against ISM properties and environment of the \thco-detected ETGs from \citet{crocker+12} and this work in addition to the spiral galaxies that are part of the 871 galaxy parent sample from which \citet{cappellari+11b} selected the \atlas\ ETGs.  From left to right: (1) $f_{60}/f_{100}$ ratio of IRAS fluxes (a dust temperature proxy); (2) M$_{\rm K}$ (a stellar mass proxy); (3) $\Sigma_3$, the 3\,Mpc galaxy density from \citet{cappellari+11} (an environment proxy).  {\bf(Bottom):} (4) $R_{\mathrm{mol}}/R_{\mathrm{e}}$, the molecular gas extent from \citet{davis+13b}.  The Spearman rank correlation coefficient (`r') and the probability that the null hypothesis (`P$_0$', no correlation) is true are given at the top of each plot.  Field ETGs are red and Virgo ETGs are blue, while spirals are green with gray error bars.  There appears to be a correlation in each with each parameter, though the galaxies within Virgo tend to be very massive and to contain more compact molecular discs. Virgo galaxies also tend to have the highest \rco\ even within individual mass bins.}
\label{fig:r13_correlations}
\end{figure*}

Figure \ref{fig:r13_correlations} isolates the ETGs and spirals in the \atlas\ parent sample \citep{cappellari+11}, performing Spearman rank statistics as was done for the original sample from \citet{crocker+12}.  The Spearman rank correlation coefficient and the probability that the null hypothesis (no correlation) is true are given. Low probabilities for the null hypothesis (we define `low' to be less than 0.03) suggest the data are correlated.  In most cases, the correlations that were originally seen in \citet{crocker+12} (including those with (1) dust temperature, (2) stellar mass and (3) environment) persist within the larger sample, and a new correlation is observed with the (4) molecular gas extent ($R_{\rm mol}/R_{\mathrm{e}}$ from \citealt{davis+13b}), which was only performed on the ETGs.  The dependence on $\Sigma_3$ (the 3\,Mpc number density; a proxy for environment from \citealt{cappellari+11}) has strengthened with the increased number statistics, but so has the dependence on galaxy mass. We are thus unable to determine whether the variation in \rco\ is independently correlated with stellar mass, environment, or molecular extent, or if our \thco-detected Virgo ETGs are simply more massive and compact than the \thco-detected field ETGs (discussed in \S\ref{sample}) or spirals, causing the correlation.

When we differentiate between types of molecular gas in ETGs, either using the CO morphological classifications from \citet{alatalo+13}, molecular gas that is disturbed morphologically preferentially has a much lower \rco\ than that of settled or aligned discs.  In fact, on average morphologically-settled molecular discs have \rco\ values almost two times higher than those of disturbed molecular distributions. When we compared the \rco\ to the kinematic misalignment between the gas and the stars ($\Psi_{mol-*}$; \citealt{davis+11}), the probability of a null hypothesis was 24\%, meaning that it appears that the disruption of the molecular gas matters far more than whether the gas was externally accreted.  Figure \ref{fig:all_co13} outlines in magenta ETGs that have been classified as ``disrupted'' by \citet{alatalo+13}, and shows that in general those galaxies have lower \rco, (in addition to high $^{12}$CO integrated flux densities, given they were part of the flux-limited dense gas survey from \citealt{crocker+12}).  

This fits quite well with the work of \citet{aalto+95}, who see similarly low \rco\ values in mergers, compared to normal spirals.  It is difficult to parse whether the differences in distributions we see are because galaxies in Virgo tend to have much more settled molecular discs, or if there are other effects at play.  Further studies of \thco\ in a large number of ETGs with settled molecular discs, in a wide variety of environments, will likely shed light on the relative importance of morphology and environment.  No previous \thco\ study includes observations of Virgo spirals.

\subsection{Possible causes of the variation in \rco, optical depth or abundance?}
In the giant molecular clouds (GMCs) of our Milky Way, typical \rco\ values are 0.2 for whole GMCs, and 0.3 for GMC centres \citep{polk+88}. So an ensemble dominated by cold GMC-like clouds should have a high \rco.  Ratios above this can be understood as arising from optically thick and very cold gas, perhaps like a GMC but with its surrounding envelope stripped away.  We know that the \thco/\twco\ line ratio correlates strongly with the dust colour (or temperature;  \citealt{young+sanders86,aalto+95}), as is clearly seen in Figure \ref{fig:all_co13}.  Alternatively, a high \rco\ value could be partially a result of enhanced $^{13}$C over $^{12}$C, although this would still require very optically thick gas. An illustration of the latter is the fact that the $^{13}$C/$^{12}$C abundance ratio decreases strongly with radius in the Milky Way \citep{milam+05}, but \rco\ has a strong opposite trend \citep{digel+90,carpenter+90,sodroski91}. This is because the \twco\ optical depth in Galactic centre clouds is significantly lower due to both temperature effects and the presence of unbound gas.

%%%%%%%%%%  Figure 7  %%%%%%%%%%

For the \thco/\twco\ line ratio, optical depth effects may either be caused by temperature/density changes (e.g. \citealt{aalto+97}), or by increased line-widths in diffuse, non self-gravitating gas (e.g. \citealt{aalto+10}).  These do not require a secondary boost in the $^{13}$C isotope compared to $^{12}$C, that has been suggested as taking place during stellar processing, specifically in low-mass stars \citep{casoli+91}.  \citet{crocker+12} considered abundance, fractionation and optical depth explanations for the trends seen in their smaller sample of ETGs, and concluded that optical depth (mediated by the gas dynamical state) dominates the \rco\ ratio changes. Here, with additional sample galaxies and more information about the gas dynamics from the interferometric maps, we return to this question. In addition, we consider a new option, the evaporation of lower-mass GMCs, or ``the survival of the densest.''

\subsubsection{Fractionation}
Carbon-bearing molecules are susceptible to a process called fractionation. This is particularly true for CO, and therefore \twco/\thco\ abundance ratios may deviate from the $^{12}$C/$^{13}$C ratio from nucleosynthesis.  There are two competing processes. The deviation may be in the form of an enhancement or a reduction depending on which process dominates.  The first process is selective photodissociation, which will enhance \twco\ since it is the more abundant isotopic variant and self-shields more effectively (e.g. \citealt{vandishoeck+black88}).  The second process operates when  gas kinetic temperatures are low to moderate (below 35 K) and elevates \thco\ through  the isotopic charge exchange reaction:

\begin{equation*}
{\rm ^{12}CO +~^{13}C^+ \longrightarrow ~^{12}C^+ + ~^{13}CO + \Delta E}.
\end{equation*}

%\[{\rm ^{12}CO + ^{13}C^+ \longrightarrow ^{12}C^+ + ^{13}CO + \Delta E} \]

\noindent This results in \thco/\twco\ abundances above the $^{13}$C/$^{12}$C ratio from nucleosynthesis by up to a factor of five (e.g. \citealt{sonnentrucker+07}), without also requiring an enhancement in the $^{13}$C/$^{12}$C abundance. 

How can we test if our \rco\ ratios are caused by ion molecule exchange reactions? The dust temperatures are low, so if they are related to the gas kinetic temperatures fractionation seems possible. \citet{ritchey+11} point out that because \twco\ is the most abundant carbon-bearing molecule in the ISM, an enhancement in \thco\ will result in a depletion of the carbon reservoir of $^{13}$C$^+$.  Other molecules originating from the remaining carbon should therefore have a $^{13}$C/$^{12}$C ratio that behaves in the opposite sense compared to \rco,  i.e. we would expect to see smaller $^{13}$C/$^{12}$C ratios in these species for Virgo galaxies.  \citet{ritchey+11} suggest CN and CH$^+$ as comparison species, and for our purposes CN should work well. CN has the additional advantage that the two adjacent spin groups can be used for an accurate measure of the optical depth. HCN and HCO$^+$ should also work quite well. We note that, for homogeneous clouds, the enhancement of \thco\ should occur mostly in the outer regions where there is sufficient $^{13}$C$^+$. If the clouds have structure the enhancement will leak further into the clouds.  %{\bf In this paper, fractionation is a viable process to have created the enhanced \rco\ that is seen.}

While fractionation is a compelling possibility to create the enhanced \rco\ ratios seen, \citet{crocker+12} did not see a correlation between $^{13}$CO/HCN (compared to $^{12}$CO/HCN) and morphological type, or temperature, potential signs that fractionation is an important driver of the \rco\ variations.  Observations of dense gas across a larger sample of molecular gas-containing ETGs would further clarify whether or not we can completely rule out fractionation, but current data suggest it is not a dominant effect.

\subsubsection{Enhanced $^{13}$C abundances from stellar reprocessing}
A possible origin for the distribution of \rco\ ratios is that the molecular gas in Virgo galaxies has resided in those galaxies for far longer than that of  average field ETGs.  \citet{davis+11} discovered that the angular momentum of ionised and molecular gas in Virgo galaxies is always aligned with the stellar angular momentum, which was posited to mean that the molecular gas within Virgo ETGs is not externally accreted, and either is due to stellar mass loss \citep{faber+gallagher76} or remains from aligned molecular gas when the galaxy first entered Virgo, thus supporting this claim.  

The observation that some Virgo spirals may have significantly higher O/H abundances within their ionised gas than field galaxies \citep{shields+91} also support this claim, although more recent metallicity studies contradict it \citep{hughes+13}. \citet{casoli+91,casoli+92a,casoli+92b} discuss low \rco\ ratios as likely being an abundance effect, with more primordial (and largely unprocessed) molecular gas containing fewer $^{13}$C atoms, since this isotope is formed via low-mass stellar nucleosynthesis.  The interstellar medium (ISM) in Virgo galaxies would have existed sufficiently long for this enrichment to occur.  NGC\,4710, which is on the outskirts of Virgo, has \rco=0.15, at the junction between most spirals, interacting galaxies, and Virgo galaxies.  If NGC\,4710 is on its first approach (suggested by its presence outside of the Virgo virial radius), and therefore just losing the ability to accrete new primordial gas, then this enrichment has just started to boost the $^{13}$C abundance and is beginning to impact \rco.

Objects with morphological signs of recent accretion events are the most likely to also have low \rco\ values, as is expected if the gas in the objects that recently accreted is more primordial and less enhanced. Though this is the case, kinematic misalignments with stars, consistent with external accretion, did not show such a distinctive difference, especially compared to the disruption seen in the gas (see \S\ref{sec:rco_var}).  If we assume that on average, the accreted molecular gas has a lower $^{13}$C abundance, then there should be a trend toward lower \rco\ from externally acquired gas, not solely morphologically disrupted gas.

Studying isotopic line ratio in other gas species (HCN vs. H$^{13}$CN, for instance) might be able to confirm whether enrichment plays a critical role in the \rco\ ratio.  Alternatively, studying \rco\ at higher excitation, and getting a different value of \rco\ in the settled discs, could rule out enrichment (though not necessarily confirm it).

\subsubsection{Midplane pressure}
Rather than being an abundance effect, the elevated \rco\ in Virgo ETGs could also be the result of midplane pressure, that can modify the optical depth of the molecular gas.  First posited by \citet{crocker+12} to describe the correlations seen in the original sample of \atlas\ ETGs, it is possible that the hot intracluster medium (ICM) within Virgo produces a larger than normal midplane pressure in galaxies.  This excess pressure on the molecular disc would compactiy it, creating a disc with higher optical depth (and thus higher \rco) than is seen in field galaxies.  The fact that the resolved maps of NGC\,4429 and NGC\,4435 do not appear to show any significant spatial variation in \rco\ seems to support the possibility of a high pressure throughout.

However, high optical depths within the discs of these galaxies would likely imply a higher fraction of dense gas, and thus impact their star formation efficiencies \citep{gao+solomon04}, but \citet{davis14} have shown that the star formation efficiencies of Virgo ETGs are similar to those of all ETGs.  Since it also appears that the molecular gas distribution plays a role in this ratio for ETGs, the high \rco\ in non-Virgo settled ETGs seems to indicate that midplane pressure is not the dominant effect in these systems.

%If we conclude that the bright \thco\ is a result of elevated \thco\ abundances then we have to consider what this means. 

%I like to call this "survival of the densest"
\subsubsection{Preferential evaporation of lower density GMCs}
It is possible that the observed enhancement of \rco\ is nothing more than ``survival of the densest'' (GMCs).  As Virgo galaxies move through the ICM, and the low density clumps are stripped (leaving only the densest clumps, likely with high \rco), one would expect our \rco\ measurements to probe only the highest end of the usual density distribution of GMCs.  \citet{pineda+08} show that the highest \rco\ values in local clumps (\rco\ $\approx$ 0.4), with derived densities of $10^3-10^4$~cm$^{-3}$, are indeed comparable to the average \rco\ of the Virgo galaxies.  The mild correlation of the HCN/\twco\ ratio with $\Sigma_3$ seen by \citet{crocker+12} also supports this hypothesis.  In fact, results from the {\em Herschel} Virgo Cluster Survey \citep{davies+12} suggest that dust and gas densities are higher in Virgo cluster galaxies than in field galaxies, likely requiring denser constituent molecular clouds.  Higher resolution studies of GMCs in Virgo galaxies with ALMA would further elucidate whether the properties of the dust transfer directly to the molecular gas properties in these systems.

\subsection{Investigating the origins of the \rco\ variation}
To fully understand the dependence of \rco\ on environment, as well as what mechanism drives that dependence, further \thco\ observations of Virgo galaxies, including Virgo spirals, will likely be required, hopefully shedding light on what is causing the large spread in \rco\ values observed (including the Virgo--field galaxy dichotomy). If $^{13}$C abundance is the most important factor, then the gas phase metallicity (a proxy for ``primordality'') should correlate with \rco.  If midplane pressure is the dominant factor for the variation in \rco, then the \rco\ ratio in varying environments (including galaxy groups and possibly the Coma Cluster) should correlate.  Finally, if entry into the hot X-ray gas halo of the Virgo cluster has stripped all but the densest clumps, then a measure of how long each galaxy has been within the cluster should correlate with \rco.  A high resolution census of the GMCs in Virgo (similar to the investigation of \citealt{utomo+15}) might also provide evidence for whether ``survival of the densest'' is able to explain the variations we see in \rco.

\section{Conclusions}
\label{conc}

The CARMA \atlas\ \twco\ survey of \citet{alatalo+13} included 17 galaxies simultaneously observed in \thco(1--0), increasing the number of \atlas\ ETGs observed  in \thco\ by 50\%, and including far fainter \twco\ detections (and more objects with settled molecular gas) than the original work of \citet{krips+10} and \citet{crocker+12}.

We presented here the \thco(1--0) resolved maps of 6 ETGs from the CARMA \atlas\ survey, those detected with sufficient significance to derive spatial-resolved \thco\ information.  An a priori masking analysis based on the method of \citet{schruba+11} allowed us to detect \thco\ in 15/17 of the observed sources, this in a spatially-integrated manner.

There does not appear to be evidence for spatial variations in \rco\ in the 6 ETGs we were able to map, though it is likely we would not have been able to detect a gradient at the level found in \citet{paglione+01}, given the signal-to-noise and resolution of our data.

Combining these new observations with those from \citet{crocker+12} and other data from the literature, we find that Virgo ETGs have a significantly higher \rco\ than both field ETGs, and spirals and interacting galaxies, although ETGs with settled molecular discs also have preferentially higher \rco\ than both spirals and ETGs with morphological disruptions.  It is also possible that the variations of \rco\ are driven by mass or molecular extent, as Virgo ETGs are, on average, more massive and contain more compact molecular discs than field ETGs.

We hypothesize that the higher \rco\ ratio in Virgo could be caused by extra low-mass stellar enrichment taking place in Virgo cluster galaxies (boosting the $^{13}$C elemental abundance in the absence of external gas accretion), by increased midplane pressure due to the ICM, or by the survival of only the densest clumps of molecular gas (as the galaxies enter the Virgo ICM).  Studies of \thco\ in more Virgo galaxies, particularly spirals, as well as studies of other $^{13}$C and $^{12}$C isotopologues, should be able to distinguish between these hypotheses.

\section*{Acknowledgments}
K.A. thanks Erik Rosolowsky, Mark Lacy, Francoise Combes, Adam Leroy, Asunci\'on Fuente and Jeff Kenney for insightful conversations that have improved this work, Theodoros Bitsakis for {\em Herschel} related questions and Selcuk Topal for a detailed investigation of the single dish data from NGC\,5866.  K.A. would also like to thank the anonymous referee for insightful comments and thoughtful recommendations.  K.A. is supported by funding through Herschel, a European Space Agency Cornerstone Mission with significant participation by NASA, through an award issued by JPL/Caltech.  M.B. is supported by the rolling grants `Astrophysics at Oxford' PP/E001114/1 and ST/H002456/1 from the UK Research Councils.  L.M.Y. is supported by the National Science Foundation under Grant No. NSF AST-1109803.
Support for CARMA construction was derived from the Gordon and Betty Moore Foundation, the Kenneth T. and Eileen L. Norris Foundation, the James S. McDonnell Foundation, the Associates of the California Institute of Technology, the University of Chicago, the states of California, Illinois, and Maryland, and the National Science Foundation. Ongoing CARMA development and operations are supported by the National Science Foundation under a cooperative agreement, and by the CARMA partner universities.  This research has made use of the NASA/IPAC Extragalactic Database (NED) which is operated by the Jet Propulsion Laboratory, California Institute of Technology, under contract with the National Aeronautics and Space Administration.

\bibliographystyle{mn2e}
\bibliography{ms}

\bsp

\label{lastpage}

\end{document}